\documentclass[12pt]{article}
\usepackage{latexsym}

\usepackage{amssymb}
\usepackage{amsmath}
\usepackage{amsfonts}
\usepackage{amsbsy}
\usepackage{amssymb}
\usepackage{graphics}
\usepackage{color}
\usepackage[colorlinks=false, pdfstartview=FitV, linkcolor=blue, citecolor=blue, urlcolor=blue]{hyperref}
\begin{document}

\noindent
{\Large{\bf Tetration: an iterative approach} }
\vskip 0.7cm
\noindent
{\bf R. Aldrovandi}
\vskip 0.2cm \noindent
{\it Instituto de F\'{\i}sica Te\'orica} \\
{\it S\~ao Paulo State University - UNESP } \\
{\it  S\~ao Paulo, Brazil}
\vskip 0.8cm
\noindent
{\bf{Abstract}}~{\em A matrix  approach to continuous iteration is proposed  for general formal series. It leads, in particular, to an order--to--order iteration of the exponential function, and consequently to an  algorithmic approach to tetration. Lower--order approximations suggest that tetration may come to be of great interest for the description of involved dynamical systems.}        


\section{Introduction: the fourth operation}
\label{sec:Introd}
%

The simplest of arithmetic operations is addition: starting with integers, $a + b $ is defined for any real or complex numbers $a$ and $b$  by an implicit interpolation process: integer $n+ m \to $ real $\to $ complex (and to more general fields).  Multiplication starts by $n a$, summing $n$ times the value $a$, and then interpolating the integers $n \to b$ to define $b a$.  Next in this hierarchy comes exponentiation: $a$ is multiplied by itself $n$ times to give $a^n$, and then $n$ is interpolated to continuum and complex values to give $a^b$.  Tetration would be next: $a \to a^a \to a^{a^a} \to ... \, a^{a^{{\cdots}^a}} $, $a$ exponentiated to $a$ $n$ times; and integer $n$ is then interpolated to real/complex values.

How to proceed to this last interpolation  ?  As $a^x = e^{(\ln a) x}$, it will be enough to consider the exponential function with basis $e$, $e^x= \exp(x)$. The formal solution would be given by solving the Schr\"oder equation~\cite{KCG90} for that exponential function, that is, by finding an invertible function $F$ such that
\begin{equation}
F[\exp(x)] = K F(x)
\label{eq:Schroder1}
\end{equation}
with $K \ne 1$ some real constant. In that case,  the n-th iterate would be $\exp^{<n>}(x) = F^{<-1>} [K^n F(x)]$ and its continuous interpolation, 
\begin{equation}
\exp^{<t>}(x) = F^{<-1>} [K^t F(x)] ,
\label{eq:formalsol}
\end{equation}
would give the $t$-th order tetration of variable $x$. The problem is that a complete solution would only come from an inspired guess of function $F$. In absence of such, we are condemned to resort to an algorithmic approach. Before going into that, let us notice that continuous iteration defines one--parameter ``families'' of functions. For example, the logarithm $\ln (x) = \ln^{<1>}(x) = \exp^{<-1>}(x)$ belongs to the same family as the exponential, as does $\exp^{<t>}(x) $ for any value of parameter $t$ and the inevitable member of every family, the identity function Id$(x) = \exp^{<0>}(x) = \ln^{<0>}(x) = x$. An easy check shows that no member on this ``exponential family'' can be the solution $F$ of Eq.(\ref{eq:Schroder1}).

A  systematic procedure to obtain  continuous  interpolations of order-by-order  iterations of certain  functions -- while keeping the sense of iteration -- have been given many years ago~\cite{AF98}.  It uses Bell matrices, and is  restricted to functions $g(x)$ whose formal Taylor series starts at order $x$: $g(x) \approx g_1 x + \frac{g_2}{2} x^2 + ... $, that is to say, to series $g(x)$ such that  $g(0) = 0$.  The main  interest lies in  dynamical systems~--~interpolations between  successive Poincar\'e sections of evolving structures~ \cite{Ald01}.  For the particular  ``extreme''   logistic map $g^{<1>}(x) = 4 x (1-x)$, for example, it gives a well-defined result.  It is enough to introduce a $\phi$ such that $x = \sin^2 \phi$ to find that $g^{<1>}(x) =  \sin^2 \left[ 2\, \arcsin \sqrt{x} \right]$, so that $g^{<k>}(x) =  \sin^2 \left[ 2^k \arcsin \sqrt{x} \,  \right]$ and $g^{<t>}(x) =  \sin^2 \left[ 2^t \arcsin \sqrt{x} \,  \right]$.

The exponential function being not of  type $g(0) = 0$,  it will require a more general approach. This will use what we shall call Carleman matrices, which generalize Bell matrices and are far more involved.  Actually, the procedure  leads to   notions much more general than tetration. 

After introducing the relationship between functions and matrices in Section~\ref{sec:Trivial}, we recall in Section \ref{sec:Bell} the method to obtain the continuous iterate of a function with vanishing independent Taylor coefficient, while profiting to introduce the standard procedure to define functions of a general non-degenerate matrix.  Section \ref{sec:Carleman} describes Carleman matrices and their main properties. Unlike Bell matrices, they have complicated eigenvalues. A generalization of the exponential is examined in Section \ref{sec:gexpon} through lower-order examples. The main problem turning up is that of convergence: in a way analogous to the  series for the exponential function,  good numerical approximations  require matrices of higher and higher orders for higher  and higher values of the arguments. 

\section{A trivial example}
\label{sec:Trivial}

 Let us begin with a  rather  trivial example: consider the two linear  functions  $G(x) = g_0 + g_1 x, \, F(y) = f_0 + f_1 y$, and to them make correspond  matrices as follows: 
\begin{gather}
G(x) = g_0 + g_1 x   \quad \longleftrightarrow \quad {\mathbb C}(G) = 
\left(  \begin{smallmatrix}
1& g_0 \\ 0 & g_1
\end{smallmatrix}  \right)\, , \label{eq:1rstexample}\\
F(x) = f_0 + f_1 x   \quad \longleftrightarrow \quad {\mathbb C}(F) = 
\left(  \begin{smallmatrix}
1& f_0 \\ 0 & f_1
\end{smallmatrix}  \right)\, .
\label{eq:2ndexample}
\end{gather}
Notice that the function coefficients can in each case be read in the second columns of the matrices.  Taking now  the function composition $[ F \circ G](x) = F[G(x)] =  f_0 + f_1 g_0 + f_1 g_1 x $, and comparing with the matrix product in inverse order (right-product), we find 
\begin{gather}
{\mathbb C}(G) {\mathbb C}(F) = \left(  \begin{smallmatrix}
1&  f_0 + f_1 g_0 \\ 0 &  f_1 g_1  
\end{smallmatrix}  \right) = {\mathbb C}(F \circ G) \, .
\label{eq:2}
\end{gather}
The composition coefficients are again read  in the second column of the resulting matrix. The relationship function $\longleftrightarrow$ matrix is consequently maintained, with  function composition corresponding to matrix right-product.  It is trivially verified that this result holds on for higher orders of composition: ${\mathbb C}(H){\mathbb C}(G) {\mathbb C}(F)  = {\mathbb C}(F \circ G \circ  H) $, etc. In particular, function iterations 
$$
G \circ G, G \circ G \circ  G,\, \ldots \, ,  \underbrace{G \circ G \circ \ldots  \circ  G \circ G}_{{\mbox{k\, times}}} ,
$$ 
for which we shall use notations $G^{<2>}$, $G^{<3>}$, $\ldots , G^{<k>}$, are represented by matrix powers: 
\begin{gather}
{\mathbb C}(G^{<2>}) =  {\mathbb C}^2(G) \, , \quad {\mathbb C}(G^{<k>}) =  {\mathbb C}^k(G) \, .
\label{}
\end{gather}

Matrix ${\mathbb C}(G)$ has  eigenvalues $\lambda_0 = 1, \lambda_1 = g_1$ and (provided $ \lambda_0 \ne \lambda_1$)  can be decomposed as
\begin{gather}
{\mathbb C}(G)  = \lambda_0 \times {\mathbb Z}_{(0)}({\mathbb C}) +  \lambda_1  \times  {\mathbb Z}_{(1)}({\mathbb C}) = 1 \times \left(  \begin{smallmatrix}
1& \frac{g_0}{1- g_1} \\ 0 & 0 
\end{smallmatrix}  \right) + g_1   \left(  \begin{smallmatrix}
0& \frac{- g_0}{1- g_1} \\ 0 & 1
\end{smallmatrix}  \right) \, .
\label{eq:Cdecomp}
\end{gather}
The two ``components''  ${\mathbb Z}_{(0)}$ and ${\mathbb Z}_{(1)}$, which appear multiplied by the corresponding eigenvalues, are  idempotent eigen-projectors ``orthogonal'' to each other: they   satisfy   ${\mathbb C} {\mathbb Z}_{(i)} = \lambda_i {\mathbb Z}_{(i)}$,  ${\mathbb Z}_{(i)} {\mathbb Z}_{(i)} =   {\mathbb Z}_{(i)}$ and, for $i \ne j$, ${\mathbb Z}_{(i)} {\mathbb Z}_{(j)} =   {\mathbb O}$ (the zero matrix). 
It is then immediate to verify that 
\begin{gather}
{\mathbb C}^2(G)  = \lambda_0^2 \times {\mathbb Z}_{(0)}({\mathbb C}) +  \lambda_1^2  \times  {\mathbb Z}_{(1)}({\mathbb C}) ;  \quad {\mbox{and}}   \\ 
{\mathbb C}^k(G)  = \lambda_0^k \times {\mathbb Z}_{(0)}({\mathbb C}) +  \lambda_1^k   \times  {\mathbb Z}_{(1)}({\mathbb C}) .
\end{gather}
This is actually a very particular example of a very general property: an arbitrary function $f$ of   matrix ${\mathbb C}(G)$ will be written 
\begin{gather}
f[{\mathbb C}(G)]  = f[\lambda_0] \times {\mathbb Z}_{(0)}({\mathbb C}) + f[ \lambda_1]   \times {\mathbb Z}_{(1)}({\mathbb C}) .
\label{eq:simpleF}
\end{gather}
This is easily proven for $f$ defined via a power series and, for non-analytic functions, is taken as the definition of $ f[{\mathbb C}(G)]$.  Notice that (\ref{eq:simpleF})  holds in this simple form only for 
 non-degenerate ${\mathbb C}(G)$, whose eigenvalues are all distinct of each other ($1 \ne g_1$ in the case).\footnote{\, The more involved approach needed for degenerate matrices can be found in references \cite{Ald01}  or \cite{Gan90}. }
 
 Continuous function iteration (say, $G^{<\alpha>}(x)$ of order $\alpha$) can then be obtained [always for the  trivial function $G(x) = g_0 + g_1 x$]  without any ado:  (i) obtain the matrix 
\begin{gather}
{\mathbb C}^\alpha(G)  = \lambda_0^\alpha \times {\mathbb Z}_{(0)}({\mathbb C}) +  \lambda_1^\alpha  
 \times {\mathbb Z}_{(1)}({\mathbb C}) =  \left(  \begin{smallmatrix}
1& \frac{g_0(1 - g_1^\alpha)}{1- g_1} \\ 0 & g_1^\alpha
\end{smallmatrix}  \right)
\end{gather}
and (ii) read the corresponding function in the second column:
\begin{gather}
G^{<\alpha>}(x) = \frac{g_0(1 - g_1^\alpha)}{1- g_1} + g_1^\alpha x \, .
\label{eq:contitera0}
\end{gather}

All this holds in this simple finite way because function $G$ is of  first order in the variable $x$~---~and so are its iterations. Higher order functions require a far more involved procedure, which is our aim to describe below.   Let us by now only retain the idea that a function can be represented by a matrix, and that in such a way that function composition is translated into matrix right--multiplication. 
 
Only functions that can be represented by Taylor series will be considered here. We shall actually use a rather loose language, interchanging expressions  ``functions" and ``formal series". And our approach will be  ``umbral'', in the sense of formal series: we shall not be concerned with convergence problems. Just as functions are in general represented by infinite series, they can also be represented by matrices. This matrix representation will actually require~---~ a drawback~---~infinite matrices. Nevertheless, finite sections of these infinite matrices will retain the character of approximations, quite analogous to that of  approximating an analytic function, equivalent to an infinite series, by a finite approximation to a certain order.

\section{Bell matrix representation}
\label{sec:Bell}

Functions of type $g(x) = \sum_{j=1}^\infty g_j \frac{x^j}{j!}$,  given by series with no independent term, are linearly represented by their Bell matrices ${\mathbb B}[g]$, whose entries ${\mathbb B}_{nm}[g]$ are provided by the multinomial theorem~\cite{Com74}
\begin{gather}
\frac{1}{m!} \left[  \sum_{j=1}^\infty g_j \frac{x^j}{j!}  \right]^m = \sum_{n=m}^\infty \frac{x^n}{n!} \,\, {\mathbb B}_{nm}[g]  \, .
\label{eq:MulttTheor}
\end{gather}
They are, thus,  just the Taylor coefficients  
\begin{gather}
{\mathbb B}_{nm}[g]  = \frac{1}{m!} \left[ 
\frac{d^{n} \ }{dx^{n}} \left( \sum^\infty_{j=1} \frac{g_j}{j!}\ x^j 
\right)^{m} \right]_{x=0} = \left[\frac {[g(x)]^m} {m!} \right]_{n}  .   
\label{eq:asTaylors}
\end{gather}
Bell matrices are lower--triangular, as ${\mathbb B}_{nm}[g] =0$ for $m> n$. The main diagonal exhibits just their eigenvalues $\{\lambda_j\} = (g_1, g_1^2, g_1^3, \ldots )$. 
Actually, as said above, complete representations of even the simplest functions would require matrices of infinite order. Nevertheless, $N \times N$ matrices retain all properties of each series up to order $N$, and can be seen as approximations to that order.  The Bell matrix will, in that case,  have the form
\begin{equation} %
{\mathbb B}[g] = \left( \begin{array}{ccccccc} g_1 & 0 & 0 & 0 & 
\cdots & \cdots & 0 \\
g_2 & g_1^2 & 0 & 0 &\cdots & \cdots & 0 \\
g_3 & 3 g_1 g_2 & g_1^3 & 0 &\cdots & \cdots & 0 \\
g_4 & 4 g_1 g_3 + 3 g_2^2 & 6 g_1^2 g_2 & g_1^4 &\cdots & \cdots & 0 
\\
\cdots & \cdots & \cdots & \cdots & \cdots & \cdots & 0 \\
g_N & \cdots & \cdots & \cdots & \cdots & \cdots & g_1^N 
\end{array} \right)   .  \label{matrix}
\end{equation} %
Notice that series $g(x)$ can be ``read'' directly from the first column:
\begin{gather}
g(x) =  \sum_{k=1}^\infty {\mathbb B}_{k 1} \frac{x^k}{k!}   \, .
\label{eq:Btoalpha}
\end{gather} 

The Bell matrix representation is of special interest for function iterations~\cite{Ald01}. Given two functions  $f(x) = \sum_{j=1}^\infty f_j \frac{x^j}{j!}$ and $g(x) = \sum_{j=1}^\infty g_j \frac{x^j}{j!}$, simple substitution shows that their composition $\; [f \circ g](x) = f[g(x)] \; $ is represented by the corresponding matrix product, though in inverse order:
\begin{equation} %
{\mathbb B}[ f \circ g]   \ = \  {\mathbb B}[g] \ {\mathbb B}[f] .  
\label{comprod} %
\end{equation} %
Function composition is, in this way, translated into matrix right-product. This means, in particular, that simple  iterations of a function are given by matrix powers:  $ {\mathbb B}[ g\circ g]   =  {\mathbb B}[ g^{<2>}]   = {\mathbb B}^2[g] $. If we use notation $g^{<k>}(x)$ for the $k-$th iterate of function $g$, then  $ {\mathbb B}[ g^{<k>}]   = {\mathbb B}^k[g] $. This can be consistently extended to iterations of  any real or complex order 
\begin{gather}
 {\mathbb B}[g^{<\alpha>}]   = {\mathbb B}^{\alpha}[g] \, ,
\label{eq:Btoanyorder}
\end{gather}
provided a meaning can be given to matrix function ${\mathbb B}^{\alpha}$.  

Whenever ${\mathbb B}[g]$ is non-degenerate (that is, if all eigenvalues are different, or simply $g_1 \ne 1$),\footnote{\, Case  $g_1 = 1$ can be dealt with separately~\cite{Ald01}. } the standard procedure to obtain a (power-series--defined) function $F({\mathbb B})$ of  matrix ${\mathbb B}[g]$  consists of the following steps: 
\begin{itemize}
\item  finding  its component--projectors ${\mathbb Z}_{(j)}[{\mathbb B}]$ such that
\begin{gather}
{\mathbb B}[g]\,  {\mathbb Z}_{(j)}[{\mathbb B}] =  \lambda_j\, {\mathbb Z}_{(j)}[{\mathbb B}] \, ,\,\,\,   \sum_{j=1}^N   {\mathbb Z}_{(j)}[{\mathbb B}]  = {\mathbb I} ; \\
{\mbox{tr}} \,   {\mathbb Z}_{(j)}[{\mathbb B}] =  1 \, , \,\,\, 
{\mathbb Z}^2_{(j)}[{\mathbb B}] = {\mathbb Z}_{(j)}[{\mathbb B}] \, , \,\,\,  {\mathbb Z}_{(i\ne j)}[{\mathbb B}] \,{\mathbb Z}_{(j)}[{\mathbb B}] = {\mathbb O}   ;
\label{eq:onZofB}
\end{gather}

\item   decomposing ${\mathbb B}[g]$  in terms of the  corresponding eigenvalues $\lambda_j = g_1^j$,
\begin{gather}
 {\mathbb B} = \sum_{j=1}^N \lambda_j \,\, {\mathbb Z}_{(j)}[{\mathbb B}]  \, ;
\label{eq:Bdecomp}
\end{gather} 

\item  writing  function $F({\mathbb B})$ as the matrix
\begin{gather}
F({\mathbb B})  = \sum_{j=1}^N F(\lambda_j )\,\, {\mathbb Z}_{(j)}[{\mathbb B}] = \sum_{j=1}^N F(g_1^j) \,\, {\mathbb Z}_{(j)}[{\mathbb B}]  \, .
\label{eq:BFunction}
\end{gather} 
\end{itemize}
In particular, 
\begin{gather}
 {\mathbb B}^{\alpha} = \sum_{j=1}^N g_1^{j \alpha}\,\, {\mathbb Z}_{(j)}[{\mathbb B}]  \, .
\label{eq:Btoalpha}
\end{gather} 
Series $g^{<\alpha>}$ can then be  ``read''  in the first column of $ {\mathbb B}^{\alpha}$:
\begin{gather}
g^{<\alpha>}(x) =  \sum_{k=1}^\infty {\mathbb B}^{\alpha}_{k 1}\, \frac{x^k}{k!}  =  \sum_{k=1}^\infty \left[ \sum_{j=1}^N g_1^{j \alpha} {\mathbb Z}_{(j) k 1}[{\mathbb B}]\right] \, \frac{x^k}{k!}   \,\,  .
\label{eq:galpha}
\end{gather} 
In other words, the Taylor coefficients of  $g^{<\alpha>}(x)  $ are
\begin{gather}
g^{<\alpha>}_k =  \sum_{j=1}^N g_1^{j \alpha}\; {\mathbb Z}_{(j) k 1}[{\mathbb B}]  \, .
\label{eq:galphak}
\end{gather}
Not every function of  ${\mathbb B}[g]$ is the Bell matrix of some function. Powers ${\mathbb B}^{<\alpha>}[g]$ are, and represent iterations $g^{<\alpha>}(x)$ of the original function.  For general values of  $\alpha$, they provide a meaning for the notion of  continuous (or more general) iteration, as they respect the (semi-)group conditions necessary for that~\cite{AF98}. 
Of particular interest  is the inverse series, obtained for $\alpha = - 1$ whenever $g_1 \ne 0$:
\begin{gather}
g^{<- 1>}(x) =  \sum_{k=1}^\infty {\mathbb B}^{- 1}_{k 1}\, \frac{x^k}{k!}  =  \sum_{k=1}^\infty \left[ \sum_{j=1}^N g_1^{- j } {\mathbb Z}_{(j) k 1}[{\mathbb B}]\right] \, \frac{x^k}{k!}   \,\,  .
\label{eq:galphaminus}
\end{gather} 
This gives just the classical Lagrange formula for series inversion~\cite{Ald01}.   

The main shortcoming  of the method is clear: even a finite polynomial will be represented by an infinite matrix.  An alternative, functional  approach which avoids all this matrix procedure has been mentioned above: given a function $g(x)$,  suppose we are happy enough to find another function $f(y)$ which solves the Schr\"oder equation 
\begin{gather}
[f \circ g] (x) = f[g(x)] = K f(x),
\label{eq:Schroder}
\end{gather}
with $K$ some constant $\neq 1$. Then, if  $f$ is invertible, $g(x) = f^{<-1>}[K f(x)]$ and the consequent $g^{<j>}(x) = f^{<-1>}[K^j f(x)]$ can be interpolated to $g^{<\alpha>}(x)$ with the same arguments given above. The difficulty lies precisely  in ``guessing'' function $f$.   An example in which guessing is not difficult  is the simplified equation for Smoluchowski's coagulation process~\cite{Cha43}, which  gives for the number of coalesced particles at time $t$ the expression
 \begin{gather}
g^{<t>}(x)    = \frac{x}{1+ x\, t}  \,\, .
\end{gather}
The Schr\"oder solution $F$ for $g^{<1>}(x)$, such that
\begin{gather}
F\left[\frac{x}{1+ x }\right] = e^{\omega} F[x]\,\, ,
\label{eq:genSchr1}
\end{gather}
is $F[u] = e^{\frac{u+\omega}{u}}$, with $F^{<-1>}[v] = \frac{\omega}{\ln v - 1}$. Then, 
 \begin{gather}
\frac{x}{1+ x }  = F^{<-1>}[ e^{\omega} F[x]] ,
\end{gather}
for any  $\omega \ne 0$. 

For functions with vanishing independent term, equation (\ref{eq:Schroder}) can be transcribed into the matrix approach as
\begin{gather}
 \sum_{r=s}^k  {\mathbb B}_{k r}[g]\; {\mathbb B}_{rs}[f]  = {\mathbb B}_{ks}[ K f] = K^s \; {\mathbb B}_{ks}[f] \label{SchrodinMat1} ;   \quad \mbox{in particular,} \quad \\
  \sum_{r=1}^k  {\mathbb B}_{k r}[g]\; {\mathbb B}_{r1}[f]  = \sum_{r=1}^k  {\mathbb B}_{k r}[g]\; f_{r} =   K \;{\mathbb B}_{k1}[f] = K\, f_{k} \, .
\label{SchrodinMat2}
\end{gather}
This eigenvalue problem allows a step-by-step calculation, but seldom leads to a closed result.

\section{Carleman matrices}
\label{sec:Carleman}

We shall use this name for extensions of the above   matrices to 
functions of type $G(x) = g_ 0 + g(x) = g_ 0 + \sum _ {j = 1}^{\infty}\frac {g_j} {j!} x^j$, given by series including an independent term.  If we define, in a way analogous to Eq.(\ref{eq:MulttTheor}),  Carleman matrices $\mathbb {C}$ with entries  $\mathbb {C}_{n r}[G]$  given by
\begin{gather}
\frac {1} {r!}\, [G(x)]^r =   \sum _ {n = 0}^{\infty}\frac {x^n} {n!}\, \mathbb {C}_{n r}[G], 
\label{eq:Cdef}
\end{gather}
a direct calculation shows that 
\begin{gather}
\mathbb {C} _ {\text {nr}}[
    G] =   \sum _ {m = 0}^{\min (r, n)}\frac {g_ 
     0 {}^{r - m}} {(r - m)!}  \left[\frac {[g(x)]^m} {m!} \right]_{_n}  = \sum _ {m = 0}^{\min (r, n)}\frac {g_ 
    0 {}^{r - m}} {(r - m)!}\;  \mathbb {B}_ {n m}[g] \, ,
\end{gather}
where the Bell matrix of  $g(x)= G(x) - g_ 0$ turns up.  With the convention $\mathbb {B}_{N 0}[g] = \delta_{N 0}$, a practical way to work with Carleman matrices\footnote{\, Notice that this name is  frequently used for the transpose  of ${\mathbb C}$ in the literature.} is the following: 
\begin{enumerate}
\item  enlarge the Bell matrix  $ {\mathbb B}[g] $  to a matrix  $ {\hat{\mathbb B}}[g] $, by adding an extra ``zero-th'' row  and an extra ``zero-th'' column, with entries $ {\hat{\mathbb B}}_{0,k\ne 0}[g] =  {\hat{\mathbb B}}_{k\ne 0,0}[g] = 0$ and $ {\hat{\mathbb B}}_{00}[g] = 1$; it will get  the aspect
\begin{gather}
{\hat{\mathbb B}}[g] = \left(
\begin{array}{ll}
 1 & 0  \\
 0 & {\mathbb B}
\end{array}
\right)= 
\left(
\begin{array}{lllll}
 1 & 0 &0 & \cdots & 0 \\
 0 & {\mathbb B}_{11}[g] &0  & \cdots & 0 \\
 0 & {\mathbb B}_{21}[g] & {\mathbb B}_{22}[g]   & \cdots & 0\\
 \cdots  & {\mathbb B}_{31}[g] & {\mathbb B}_{3 2}[g]   & \cdots & 0\\
0  & {\mathbb B}_{N1}[g] & {\mathbb B}_{N 2}[g]   & \cdots &  {\mathbb B}_{N N}[g] 
\end{array}
\right)\, ; 
\label{enlargedB}
\end{gather}

\item  define the upper-triangular matrix  ${\mathbb M}[g_0]$ with entries 
\begin{gather}
{\mathbb M}_{rm}[g_0] = \frac{g_0^{m-r}}{(m-r)!}\, ,
\end{gather}
whose $N = 2$ and $N = 3$ examples are
\begin{gather}
\left(
\begin{array}{lll}
 1 & g_0 & \frac{g_0^2}{2} \\
 0 & 1 & g_0 \\
 0 & 0 & 1
\end{array}
\right) \quad , \qquad  \left(
\begin{array}{llll}
 1 & g_0 & \frac{g_0^2}{2} & \frac{g_0^3}{6} \\
 0 & 1 & g_0 & \frac{g_0^2}{2} \\
 0 & 0 & 1 & g_0 \\
 0 & 0 & 0 & 1
\end{array}
\right)\,\, .
\end{gather}
${\mathbb M}[g_0]$ has properties  $\det {\mathbb M}[g_0] = 1 $  and ${\mathbb M}^n[g_0 ]= {\mathbb M} [n g_0]$.

\item  then,   
\begin{gather}
{\mathbb C} [G] = {\hat{\mathbb B}}[g]\,\, {\mathbb M}[g_0]\, , 
\label{CBM}
\end{gather}
which, by the way, shows that  $\det {\mathbb C} [G] = \det {\mathbb B}[g] $. Consequently, ${\mathbb C} [G]$ is invertible when and only when ${\mathbb B}$ (and the corresponding series) is invertible, that is, when $g_1 \ne 0$. Notice that the matrix product above includes 
the contributions of the zero-th rows and columns: ${\mathbb C}_{nm} [G] = \sum_{j =0}^N {\hat{\mathbb B}}_{nj}[g] \, {\mathbb M}_{j m} [g_0]$ are the entries of an $(N+1) \times (N+1)$ matrix.

\end{enumerate}

The $3 \times 3$ Carleman matrix, corresponding to $N=2$,  is
\begin{multline}
\left(
\begin{array}{lll}
 1 & g_0 & \frac{g_0^2}{2} \\
 0 & g_1 & g_0 g_1 \\
 0 & g_2 & g_1^2+g_0 g_2
\end{array}
\right) \\
= \left(
\begin{array}{lll}
 1 & 0 & 0\\
 0 & g_1 & 0 \\
 0 & g_2 & g_1^2
\end{array}
\right) + g_0 \left(
\begin{array}{lll}
0 & 1 & 0 \\
 0 & 0&  g_1 \\
 0 & 0 &  g_2
\end{array}
\right) + \frac{g_0^2}{2}\left(
\begin{array}{lll}
0 & 0 & 1 \\
 0 & 0 & 0 \\
 0 & 0& 0
\end{array}
\right) . 
\label{eq:C3x3}
\end{multline}

The $4 \times 4$   matrix is
\begin{multline}
\left(
\begin{array}{llll}
 1 & g_0 & \frac{g_0^2}{2} & \frac{g_0^3}{3!} \\
 0 & g_1 & g_0 g_1 & \frac{g_0^2 g_1}{2} \\
 0 & g_2 & g_1^2+g_0 g_2& g_0 g_1^2+  \frac{g_0^2 g_2}{2} \\
  0 & g_3 & 3 g_1  g_2+g_0 g_3&  g_1^3+ 3 g_0 g_1  g_2 + \frac{g_0^2 g_3}{2} 
\end{array}
\right) = \left(
\begin{array}{llll}
 1 &0 &0 &0\\
 0 & g_1 &0 & 0\\
 0 & g_2 & g_1^2& 0 \\
  0 & g_3 & 3 g_1  g_2  & g_1^3 
\end{array}
\right) +
\\
g_0 \left(
\begin{array}{llll}
0 & 1 & 0 &0\\
 0 &0 &  g_1 &0 \\
 0 & 0 & g_2&  g_1^2 \\
  0 & 0 & g_3&   3  g_1  g_2 
\end{array}
\right) + \frac{g_0^2}{2} \left(
\begin{array}{llll}
 0 & 0&1 & 0\\
 0 & 0 & 0 & g_1\\
 0 & 0 & 0& g_2 \\
  0 & 0 &  0&  g_3
\end{array}
\right) + \frac{g_0^3}{3!}\left(
\begin{array}{llll}
0 & 0&0 &1 \\
 0 &0 & 0 &0 \\
 0 & 0 & 0&0\\
  0 & 0 & 0& 0
\end{array}
\right) \, .
\end{multline}
The second column of ${\mathbb C} [G] $ can be read into the series $G(x) = g_{0} + g(x)$. Notice that, when isolating progressive  powers of $g_0$, matrix $ {\hat{\mathbb B}}[g]$ appears successively in different ways: first itself, and  then, at each order, dislocated one more step to the right. An operational version is obtained if we introduce  $(N+1) \times (N+1)$ matrices ${\mathbb W}$ whose entries are all zero, except those in the first-right diagonal, which are $= 1$. Thus, for $N = 3$, ${\mathbb W} = \left( \begin{smallmatrix} 0&1&0&0\\ 0&0&1&0\\  0 & 0 & 0&1\\
  0 & 0 & 0& 0 \end{smallmatrix} \right)$. Their right - action on another matrix causes an overall shift to the right, destroying the last column and zeroing the first.  Then, ${\mathbb M}[g_0] = \sum_{j=0}^N \frac{{\mathbb W}^j g_0^j}{j!}$ and
\begin{gather}
{\mathbb C} [G] = {\hat{\mathbb B}}[g] \exp_N[{\mathbb W} g_0]
\label{CwithExpW}
\end{gather}
  with $ \exp_N$ the exponential series truncated  to order $N$.

Now comes the important point:  these ${\mathbb C}$ maintain the relationship between composition and matrix right-product: it is indeed enough to replace \,  $x \to F(x) = f_{0} + f(x) $\,  in Eq.(\ref{eq:Cdef}) to show that, for $(N+1) \times (N+1)$ matrices, 
\begin{gather} 
{\mathbb C}_{n m}[G \circ F]  = \sum_{r=0}^{N} {\mathbb C}_{n r}[F] \,\, {\mathbb C}_{r m}[G] \, . 
\label{eq:Ccomposition}
\end{gather}
This means that we can repeat here the procedure of Section \ref{sec:Bell}.  Corresponding to Eqs.(\ref{eq:Bdecomp}-\ref{eq:galphak}), we shall now have
\begin{gather}
 {\mathbb C} = \sum_{j=0}^N \lambda_j \,\, {\mathbb Z}_{(j)}[{\mathbb C}]  \, ;  \\
F({\mathbb C})  = \sum_{j=0}^N F(\lambda_j )\,\, {\mathbb Z}_{(j)}[{\mathbb C}]  \, ; \label{eq:FofC} \\ 
 {\mathbb C}^{\alpha} = \sum_{j=0}^N \lambda_j^{\alpha}\,\, {\mathbb Z}_{(j)}[{\mathbb C}]  \, ; \\ 
G^{<\alpha>}(x) =  \sum_{k=0}^\infty {\mathbb C}^{\alpha}_{k 1}\, \frac{x^k}{k!}  =  \sum_{k=0}^\infty \left[ \sum_{j=1}^N \lambda_j^{\alpha}\, {\mathbb Z}_{(j) k 1}[{\mathbb C}]\right] \, \frac{x^k}{k!}   \,\,  ; \\
G^{<\alpha>}_k =  \sum_{j=0}^N  \lambda_j^{\alpha}\, {\mathbb Z}_{(j) k 1}[{\mathbb C}]  \, .
\label{eq:Galphak}
\end{gather}
A difficulty comes now from the eigenvalues $\lambda_j$ of ${\mathbb C}$, whose general expressions are, unlike those of ${\mathbb B}$,  non-trivial. Already for the $N = 2$ matrix (\ref{eq:C3x3}), they are $\lambda_0 = 1$ and $\lambda_{1,2} = \frac{g_1 + g_1^2+ g_0  g_2 \mp \sqrt{\left( {g_1}+ {g_1}^2+ {g_0}  {g_2}\right)^2-4 {g_1}^3}}{2} $ .  

We can alternatively define the projector--related functions
 \begin{gather}
R_j(x) =  \sum_{k=0}^\infty {\mathbb Z}_{(j) k 1}[{\mathbb C}] \, \frac{x^k}{k!}  \, ,
\label{eq:RjofC}
\end{gather}
 so that the $\alpha$-- and the $x$--dependences are, in a sense, factorized:
 \begin{gather}
G^{<\alpha>}(x) =   \sum_{j=1}^N \lambda_j^{\alpha}\, R_j(x)  \, .
\label{eq:GfromRjs}
\end{gather}

More generally, Eq.(\ref{eq:FofC}) defines a functional ${\mathcal F}[G]$ by 
\begin{gather}
{\mathcal F}[G](x) =    \sum_{k=0}^\infty \left(\sum_{j=1}^N F(\lambda_j)\, {\mathbb Z}_{(j) k 1}[{\mathbb C}] \right)\, \frac{x^k}{k!}  =  \sum_{j=1}^N F(\lambda_j)\, R_j(x)  \, ,
\label{eq:functional}
\end{gather}
or by the Taylor coefficients
\begin{gather}
{\mathcal F}[G]_k = \sum_{j=1}^N F(\lambda_j)\, {\mathbb Z}_{(j) k 1}\left[{\mathbb C}[G] \right] \, .
\label{eq:funcTaylor}
\end{gather}
These matrices actually ``over-represent'' the corresponding series at each order, as they contain a lot of redundant information. For the linear example in Section  \ref{sec:Trivial} a $2 \times 2$ matrix is enough  because  iteration of a linear function is a linear function. Instead of Eq.(\ref{eq:1rstexample}), for example,  the complete matrix would have the aspect 
\begin{gather}
\left(
\begin{smallmatrix} 
 1 & g_0 & \frac{g_0^2}{2} & \frac{g_0^3}{3!} & ... \\
 0 & g_1 & g_0 g_1 & \frac{g_0^2 g_1}{2} & ... \\
 0 & 0 & g_1^2 & g_0 g_1^2 & ...  \\
  0 & 0 &0 &  g_1^3 & ... \\
   ...& ...& ... & ... & ... \\
   ...& ...& ... & ... & ... 
\end{smallmatrix}
\right) \ \ , 
\end{gather}
but all rows and columns not present in Eq.(\ref{eq:1rstexample}) are irrelevant.   \\

\section{Example: General exponentiation}
\label{sec:gexpon}

The expressions  for the ${\mathbb C} [G]$  entries become rapidly very long with increasing $N$. It will be instructive to examine the exponential function, for which all  $g_k =1$.  For case  $N = 2$, ${\mathbb C} [e^x = 1+x+\frac{x^2}{2} + O(x^3)]$ is written in terms of its projectors and respective eigenvalues as
\begin{multline}
\left( \begin{smallmatrix}  1 & 1 & \frac{1}{2}
\\ 0&1&1\\
0&1 & 2 \end{smallmatrix} 
\right) =  
 1  \times 
\left( \begin{smallmatrix}  1& \frac{1}{2}& -1\\
0&0&0\\
0&0&0 \end{smallmatrix} 
\right) + \\
+ 
{\textstyle{ \frac{3-\sqrt{5}}{2}  }} \times
\left( \begin{smallmatrix}  0& - \frac{1+\sqrt{5}}{4}& \frac{1}{2}\\ 
0& \frac{5+\sqrt{5}}{10}& - \frac{1}{\sqrt{5}} \\
0& - \frac{1}{\sqrt{5}}& \frac{5-\sqrt{5}}{10} 
\end{smallmatrix} 
\right)+ 
{\textstyle{  \frac{3 + \sqrt{5}}{2}   }} \times
\left( \begin{smallmatrix}  0&- \frac{1-\sqrt{5}}{4}& \frac{1}{2} \\
0 & \frac{5-\sqrt{5}}{10}&\frac{1}{\sqrt{5}} \\ 0&\frac{1}{\sqrt{5}}&\frac{5+\sqrt{5}}{10}
\end{smallmatrix} 
\right). 
\label{eq:CforN2}
\end{multline}
A function $F$ of this non-degenerate matrix will be 
\begin{multline}
F\left[\left( \begin{smallmatrix}  1 & 1 & \frac{1}{2}
\\ 0&1&1\\
0&1 & 2 \end{smallmatrix} 
\right)\right] =  
 F[1]  \times 
\left( \begin{smallmatrix}  1& \frac{1}{2}& -1\\
0&0&0\\
0&0&0 \end{smallmatrix} 
\right) + \\
+ 
F\left[ {\textstyle{ \frac{3-\sqrt{5}}{2}  }} \right] \times
\left( \begin{smallmatrix}  0& - \frac{1+\sqrt{5}}{4}& \frac{1}{2}\\ 
0& \frac{5+\sqrt{5}}{10}& - \frac{1}{\sqrt{5}} \\
0& - \frac{1}{\sqrt{5}}& \frac{5-\sqrt{5}}{10} 
\end{smallmatrix} 
\right)+ 
F\left[{\textstyle{  \frac{3 + \sqrt{5}}{2}   }}  \right] \times
\left( \begin{smallmatrix}  0&- \frac{1-\sqrt{5}}{4}& \frac{1}{2} \\
0 & \frac{5-\sqrt{5}}{10}&\frac{1}{\sqrt{5}} \\ 0&\frac{1}{\sqrt{5}}&\frac{5+\sqrt{5}}{10}
\end{smallmatrix} 
\right). 
\label{eq:FforN2}
\end{multline}
Examples are 
\begin{description}
\item[(i)]  the identity matrix, which comes for $F(u) = u^0 = 1$, and whose 2nd column gives just $G(x) = x$;

\item[(ii)] the identity function $F(u) = u^1 = u$, leading to $F\left[ \left( \begin{smallmatrix}  1 & 1 & \frac{1}{2}
\\ 0&1&1\\
0&1 & 2 \end{smallmatrix} 
\right)  \right] = \left( \begin{smallmatrix}  1 & 1 & \frac{1}{2}
\\ 0&1&1\\
0&1 & 2 \end{smallmatrix} 
\right) $, whose  2nd column gives $G(x) = 1+x+\frac{x^2}{2}$; 

\item[(iii)]  and the arbitrary power
\begin{multline}
 \left( \begin{smallmatrix}  1 & 1 & \frac{1}{2}
\\ 0&1&1\\
0&1 & 2 \end{smallmatrix} 
\right)^t =  
\left( \begin{smallmatrix}  1& \frac{1}{2}& -1\\
0&0&0\\
0&0&0 \end{smallmatrix} 
\right) + \\
+ 
 \left[{\textstyle{\frac{3 - \sqrt{5}}{2} }} \right]^t  \times
\left( \begin{smallmatrix}  0& - \frac{1+\sqrt{5}}{4}& \frac{1}{2}\\ 
0& \frac{5+\sqrt{5}}{10}& - \frac{1}{\sqrt{5}} \\
0& - \frac{1}{\sqrt{5}}& \frac{5-\sqrt{5}}{10} 
\end{smallmatrix} 
\right) + 
 \left[{\textstyle{\frac{3 + \sqrt{5}}{2} }} \right]^t \times
\left( \begin{smallmatrix}  0&- \frac{1-\sqrt{5}}{4}& \frac{1}{2} \\
0 & \frac{5-\sqrt{5}}{10}&\frac{1}{\sqrt{5}} \\ 0&\frac{1}{\sqrt{5}}&\frac{5+\sqrt{5}}{10}
\end{smallmatrix} 
\right). 
\label{eq:Ctot}
\end{multline}
\end{description}
 It is easily verified that $t = -1$ does provide the inverse to matrix (\ref{eq:CforN2}). By the way,  this illustrates also a statement made above: the procedure is self-consistent at each order. At the $N$-th level,   the inverse matrix found  is just  the inverse to the Carleman $(N+1) \times (N+1)$ matrix. 
Reading along the 2nd columns in (\ref{eq:Ctot}), 
\begin{multline}
G^{<t>}(x) = {\textstyle{ \frac{1}{2} -  \frac{1-\sqrt{5}}{4} \left(\frac{3+\sqrt{5}}{2}\right)^t   - \frac{1+\sqrt{5}}{4}\left(\frac{3-\sqrt{5}}{2}\right)^t  }}  \\
+ {\textstyle{  \left[\frac{5-\sqrt{5}}{10} \left(\frac{3+\sqrt{5}}{2}\right)^t+ \frac{5+\sqrt{5}}{10}\left(\frac{3-\sqrt{5}}{2}\right)^t\right] }} x    \\
+ {\textstyle{ \frac{1}{\sqrt{5}} \left[\left(\frac{3+\sqrt{5}}{2}\right)^t-\left(\frac{3-\sqrt{5}}{2}\right)^t \right]   }}\frac{x^2}{2}\, ,
\label{eq:tetN2}
\end{multline}
or  
\begin{multline}
G^{<t>}(x) = {\textstyle{\frac{1}{2} + \left(\frac{3+\sqrt{5}}{2}\right)^t  \left[ \frac{5-\sqrt{5}}{10} \, x -  \frac{1-\sqrt{5}}{4} + \frac{1}{\sqrt{5}} \frac{x^2}{2} \right]  }} \\
  + {\textstyle{ \left(\frac{3-\sqrt{5}}{2}\right)^t  \left[ \frac{5+\sqrt{5}}{10} x - \frac{1+\sqrt{5}}{4}- \frac{1}{\sqrt{5}} \frac{x^2}{2}  \right]  }} .
\label{eq:tetN2other}
\end{multline}
For integer values of $t$, this gives the approximate  expressions: $G^{<0>}(x) = x$, $G^{<1>}(x) = 1+x+\frac{x^2}{2}$,  $G^{<2 >}(x) = \frac{1}{2} \left(5+4 x+3 x^2\right) = 1 + G^{<1>}(x)+ \frac{1}{2} (G^{<1>}(x))^2$.


Equations (\ref{eq:tetN2}, \ref{eq:tetN2other}) give the  tetration values for function $e^x$ to order $N=2$.  If we want $e$ exponentiated by $e$ itself  $t$ times,  it is enough to put $x = e$. And to have $e$ exponentiated by $e$ itself  ``$e$ times'',  it is enough to put also $t = e$. Approximations to order $N = 2$ are, in this case, very  poor: we find $G^{<1>}(e) = 7.41281$, to be compared to $e^e \approx 15.1543$, Thus, the  found  value $G^{<e>}(e) =  37.5795$ is not trustworthy.  But this is consistent with the exponential function itself:   to   evaluate $e^e$ with less than  $1 \%$ error,  order $N = 7 $ would be necessary.  



It is fascinating that expression (\ref{eq:tetN2other}) can be rewritten in terms of the Golden Ratio $\alpha = \frac{1+\sqrt{5}}{2}$:
\begin{multline}
G^{<t>}(x) = {\textstyle{\frac{1}{2} + \left(1+ \alpha\right)^t  \left[ \frac{3- \alpha}{5} \, x -  \frac{1- \alpha}{2} +   \frac{x^2}{4  \alpha - 2} \right]  }} \\
  + {\textstyle{ \left(2 -  \alpha \right)^t  \left[ \frac{2+ \alpha}{5}\; x -  \frac{ \alpha}{2} -  \frac{x^2}{4  \alpha - 2} \right]  }} \ . 
  \label{eq:tetN2Goldie}
  \end{multline}

Of course, in terms of  Carleman matrices, solving the  Schr\"oder equation $\mathbb {C}[g]\mathbb {C}[F] = \mathbb {C}[K F] $ becomes finding eigenvectors: as $\mathbb {C}_{n1}[f] = f_n$, it becomes
\begin{gather}
\sum_{s=0}^{\infty} \mathbb {C}_{rs}[g] \mathbb {C}_{s1}[F] = \mathbb {C}_{r1}[K F] ,  \quad {\mbox{which is}} \\
\sum_{s=0}^{\infty} \mathbb{C}_{r s}[g] F_{s} = K F_r  \,\,  .
\end{gather}
In particular, for tetration we should find the general eigenvectors  of 
\begin{gather}
 \mathbb {C}_{nj}[e^x] = \frac{j^n}{j!} \, \, .
\end{gather}

A curious point is that the sum of all such matrix elements in a line is related to Stirling numbers. A hallmark property of the Stirling numbers~\cite{AS68} of the second kind $S_{n}^{(k)}$ is the expression
\begin{equation}
\left(x \; \frac{d\ }{dx}\right)^n f(x) = \sum_{k=1}^{n} S_{n}^{(k)} x^k \frac{d^k\ }{dx^k}
 f(x)  \,  \label{useful1}
\end{equation}
which gives, when applied to the exponential,

\begin{equation}
\sum_{j=0}^{\infty} \frac{j^{n}}{j!}\; x^j = \left(x \;  \frac{d\ }{dx}\right)^n e^x  =  e^x  \sum_{k=1}^{n} S_{n}^{(k)} x^k \  . \label{dnexp }
\end{equation}
This  leads,  for $x = 1$, to  the Dobi\'nski formula
\begin{equation}
\sum_{j=0}^{\infty} \frac{j^{n}}{j!}\  = e \sum_{k=1}^{n} S_{n}^{(k)}   = e\  \omega(n) 
 . \label{dobinski}
\end{equation}
The Bell number $\omega(n) =  \sum_{k=1}^{n} S_{n}^{(k)}  $ is the number of ways of partitioning 
a set of $n$ members. Taking by convention  $\omega(0) = 1 $, we have that 
\begin{equation}
\sum_{j=0}^{\infty}  \mathbb {C}_{nj}[e^x] =  e\  \omega(n)  ,  \label{exp2}
\end{equation}
the sum of all the elements in the $n$-th row of $\mathbb {C}_{nj}[e^x] $. For the $0$-th row it gives, of course, just $\exp(1)$. From \ref{eq:Cdef}, this says just that 
\begin{gather}
G^{<2>}(x) = e^{e^x} = e \sum_{j=0}^{\infty}  \omega(n)\;   \frac{x^n}{n!} \  \  .
\label{}
\end{gather}

\section{Final comments}
\label{sec:fincomm}

Continuous iteration is, of course, a much more general subject. We have here restricted ourselves to its application to  tetration  \cite{Forum,Trappmann}.  It would be desirable to  examine the differential equations which would come out of its use.  Continuous iteration involves a ``non-locality'', a global feature in function space. Functions constitute families. The  ``exponential family" of functions $\exp^{<\alpha>}(x) = \ln^{<- \alpha>}(x)  $ will include the exponential $\exp (x) = \exp^{<1>}(x)$, the logarithm $\ln(x) = \exp^{<-1>}(x)$ and the inevitable member of every family, the identity function Id$(x) = \exp^{<0>}(x) = \ln^{<0>}(x) = x$. Let us remember that many physical phenomena, from the decay of subatomic particles  to the cosmological expansion of the Universe, could not be described before the introduction of exponentiation  in the XVIII-th century.  It is to be hoped that this "fourth operation" come to be of interest to the description of some as yet untamed phenomena. The appearance of Stirling numbers~---and the alluring presence of the Golden Ratio in the first approximations~---~is an intimation of  relationship to  complex evolving systems.

\section*{Aknowledgements}
The author is particularly grateful to Daniel Geisler~\cite{Geisler}  and to Gottfried Helms~\cite{Helms} for indirect and direct encouragement. 
A text on Carleman matrices, including an example with a Mathematica program,  has been posted by Gottfried Helms, and  can be retrieved from:

http://math.eretrandre.org/tetrationforum/showthread.php?tid=149

%
%
%
%
%
%
%


\end{document}